\newif\ifmsinclaude
\makeatletter
\IfFileExists{../references.bib}{\msinclaudetrue\def\input@path{{./}{../}}}{\def\input@path{{./}{Claude_v2/}}}
\makeatother
\newif\ifmsbibinclaude
\ifmsinclaude
  \edef\msjobname{\jobname}\edef\msoverleafjob{\detokenize{output}}
  \ifx\msjobname\msoverleafjob\else\msbibinclaudetrue\fi
\fi
\documentclass[fleqn,usenatbib]{mnras}
\usepackage{newtxtext,newtxmath}

\usepackage[T1]{fontenc}

\DeclareRobustCommand{\VAN}[3]{#2}
\let\VANthebibliography\thebibliography
\def\thebibliography{\DeclareRobustCommand{\VAN}[3]{##3}\VANthebibliography}

\usepackage{graphicx}	
\usepackage{amsmath}	
\usepackage{comment}
\usepackage{xcolor}	
\usepackage{scalerel,tikz}
\usetikzlibrary{svg.path}
\definecolor{orcidlogocol}{HTML}{A6CE39}
\tikzset{orcidlogo/.pic={
    \fill[orcidlogocol] svg{M256,128c0,70.7-57.3,128-128,128C57.3,256,0,198.7,0,128C0,57.3,57.3,0,128,0C198.7,0,256,57.3,256,128z};
    \fill[white] svg{M86.3,186.2H70.9V79.1h15.4v48.4V186.2z}
    svg{M108.9,79.1h41.6c39.6,0,57,28.3,57,53.6c0,27.5-21.5,53.6-56.8,53.6h-41.8V79.1z M124.3,172.4h24.5c34.9,0,42.9-26.5,42.9-39.7c0-21.5-13.7-39.7-43.7-39.7h-23.7V172.4z}
    svg{M88.7,56.8c0,5.5-4.5,10.1-10.1,10.1c-5.6,0-10.1-4.6-10.1-10.1c0-5.6,4.5-10.1,10.1-10.1C84.2,46.7,88.7,51.3,88.7,56.8z};
}}
\newcommand\orcidicon[1]{\href{https://orcid.org/#1}{\mbox{\scalerel*{
    \begin{tikzpicture}[yscale=-1,transform shape]
        \pic{orcidlogo};
    \end{tikzpicture}
}{|}}}}
\usepackage{hyperref}

\title[Ly$\alpha$-guided 21-cm troughs]{A tale of two forests -- I. Finding broad 21-cm forest troughs with a Ly$\alpha$-informed matched filter}

\author[Qin et al.]{Yuxiang Qin~\orcidicon{0000-0002-4314-1810}$^{1}$\thanks{E-mail: Yuxiang.L.Qin@gmail.com},
Andrei Mesinger~\orcidicon{0000-0003-3374-1772}$^{2}$,
Gianni~Bernardi~\orcidicon{0000-0002-0916-7443}$^{3,4,5}$, 
Valentina D'Odorico~\orcidicon{0000-0003-3693-3091}$^{6,7,8}$, \newauthor 
Chanasorn~Kongprachaya$^{3,9}$ and Tomáš Šoltinský~\orcidicon{0000-0001-7703-8929}$^{6,8}$
\\
$^{1}$Research School of Astronomy and Astrophysics, Australian National University, Canberra, ACT 2611, Australia\\
$^{2}$Department of Physics and Astronomy ``Ettore Majorana'', University of Catania, Via S. Sofia 64, 95123 Catania, Italy\\
$^{3}$INAF -- Istituto di Radioastronomia, Via P.~Gobetti 101, 40129 Bologna, Italy \\
$^{4}$Centre for Radio Astronomy Techniques and Technologies (RATT), Department of Physics and Electronics, Rhodes University, Makhanda 6140, South Africa \\
$^{5}$South African Radio Astronomy Observatory, Cape Town 7700, South Africa \\
$^{6}$INAF-Osservatorio Astronomico di Trieste, Via Tiepolo 11, I-34143 Trieste, Italy\\
$^{7}$Scuola Normale Superiore, Piazza dei Cavalieri 7, I-56126 Pisa, Italy\\
$^{8}$IFPU-Institute for Fundamental Physics of the Universe, via Beirut 2, I-34151 Trieste, Italy\\
$^{9}$Department of Physics and Astronomy, University of Bologna, Via Gobetti 93/2 – 40129 Bologna, Italy
}

\pubyear{\the\year{}}

\begin{document}
\label{firstpage}
\pagerange{\pageref{firstpage}--\pageref{lastpage}}
\maketitle

\begin{abstract}
Increasing evidence for late reionization suggests that extended ``islands'' of neutral intergalactic medium persist below $z{\sim}6$. These islands can imprint broad 21-cm forest troughs towards radio-bright sources, but the shallow features are easily masked or mimicked by noise and systematic residuals.
Here, we demonstrate how to robustly identify such troughs combining (i) a multiscale matched filter, which efficiently identifies localized features in noisy data; and (ii) a mask using the Ly$\alpha$ forest, which penalizes spurious detections by requiring genuine radio troughs are also dark in Ly$\alpha$.
We showcase the performance of our method on mock spectra varying radio noise, systematic residuals, gas spin temperature, and optical signal-to-noise in the Ly$\alpha$ forest.
Without heating and systematics, the radio-only search recovers 2/3 of the troughs at 0.1\,mJy channel noise, roughly corresponding to an 18-hr observation with the upgraded Giant Metrewave Radio Telescope. The Ly$\alpha$ mask raises the purity from 82\% to 93\% with almost no loss of completeness. Sinusoidal residuals with periods of a few MHz create up to ${{\sim}}10$ false positives per spectrum at the thermal-noise level. The mask halves this contamination, outperforming the common polynomial continuum fitting that either leaves more false positives or removes additional genuine troughs.
Heating leaves almost no troughs deeper than 5 per cent. Recovering shallower troughs at 10--20\,K requires a thermal noise of 0.05--0.1\,mJy, achievable with ${\lesssim}$10-hr integration using the Square Kilometre Array.
 Our results illustrate a promising new avenue of using multi-wavelength information to map reionization morphology at late stages.
\end{abstract}




\section{Introduction}
\label{sec:introduction}

The Epoch of Reionization (EoR) marks the transformation of the intergalactic medium (IGM) from a predominantly neutral to an ionized state. Its timing and morphology encode the properties of the first galaxies and their ionizing radiation. Recent analyses of high-redshift Ly$\alpha$ transmission and quasar damping wings increasingly favour a late end to reionization, with extended neutral islands possibly surviving below $z=6$ \citep{Zhu2021ApJ...923..223Z,Bosman2022MNRAS.514...55B,Greig2024MNRAS.530.3208G,Qin2025PASA...42...49Q}. 
The sizes and properties (e.g. temperature and density) of these cosmic neutral patches encode invaluable information on IGM clumping (e.g. \citealt{Choudhury2009MNRAS.394..960C,Sobacchi2014MNRAS.440.1662S}), as well as the ionizing and X-ray properties of the first galaxies (e.g. \citealt{Greig2017MNRAS.472.2651G,Park2019MNRAS.484..933P}).
If such islands survive to these lower redshifts, their 21-cm signal could be within reach of current radio facilities, before the next generation of interferometers comes online \citep{HERA2023ApJ...945..124H,Nunhokee2025ApJ...989...57N}.

In addition to the 21-cm signal measured against the cosmic microwave background,
neutral hydrogen absorbs the emission of radio-bright background sources, producing the 21-cm forest \citep{Carilli2002ApJ...577...22C,Ciardi2003ApJ...596....1C,Mack2012MNRAS.425.2988M}. Compact absorbers such as minihalos or filaments \citep{Furlanetto2002ApJ...579....1F,Semelin2016MNRAS.455..962S,Naruse2024JCAP...10..091N} generate narrow features, whereas extended neutral islands in the IGM can imprint broad absorption troughs spanning multiple frequency channels \citep{Soltinsky2023MNRAS.519.3027S,Shimabukuro2023PhRvD.107l3520S}. A late EoR makes these broad troughs more promising targets, because neutral gas then survives to redshifts at which sufficiently bright radio sources could be more common than at earlier times \citep{Kongprachaya2026arXiv260720371K,Soltinsky2021MNRAS.506.5818S,Soltinsky2026arXiv260715341S}. However, these troughs are hard to find. Their depths can be comparable to the channel noise, and their widths could be comparable to the frequency scales of residual systematics. Moreover, their depth scales inversely with the spin temperature of the neutral gas, which is poorly constrained at these redshifts \citep{HERA2023ApJ...945..124H,Nunhokee2025ApJ...989...57N}.

Rather than identifying individual troughs, most current 21-cm forest analyses target statistical summaries such as the 1D power spectrum \citep[e.g.][]{Ewall-Wice2014MNRAS.441.2476E,Thyagarajan2020ApJ...899...16T,Soltinsky2025MNRAS.537..364S,Shao2023NatAs...7.1116S,Zhao2026PhRvD.113d3531Z}, wavelet scattering transforms \citep{Shimabukuro2025PhRvD.112f3557S} and topological signatures \citep{Shimabukuro2026PhRvD.113h3525S}.  Furthermore, implicit likelihood and other machine-learning studies have demonstrated good parameter inference from idealized mock 21-cm forest spectra \citep{Sun2025ComPh...8..220S,Patil2026MNRAS.546.stag236P}; unfortunately, these do not produce an interpretable catalogue of individual trough candidates, or validate such candidates against independent data.  More fundamentally however, most of these summaries are either not optimal for spectrally-localized features such as those imprinted by neutral islands, or are brittle in dealing with noise and residual systematics.

In this work we introduce matched filters to 21-cm forest analysis.  Matched filters are commonly used to identify localized features of known shapes from noisy data \citep{Datta2007MNRAS.382..809D,Loomis2018AJ....155..182L}.  They do so by convolving the noisy data with a set of template shapes for the sought-after feature, maximizing the resulting signal-to-noise ratio.  In this way, a matched filter offers a transparent way to identify localized absorption features of unknown width.

However, the radio spectrum alone cannot tell a genuine trough from an instrumental residual of the same shape. The neutral islands that produce broad 21-cm troughs should also suppress Ly$\alpha$ transmission along the same line of sight. A 21-cm trough should therefore coincide with an extended Ly$\alpha$-dark region at the same redshift, so the optical spectrum can independently validate candidates from the radio search. 

Here we demonstrate how a co-spatial Ly$\alpha$ spectrum can mitigate residual systematics in radio data by reducing the rate of false positives.  We filter the radio spectrum over the expected range of trough widths, and rebin the noisy optical spectrum onto the radio channels to build a Ly$\alpha$-dark mask whose threshold is set by the optical noise. The mask rejects radio candidates without Ly$\alpha$ absorption, and we score the resulting catalogue by its completeness and purity, matching each detection to at most one true trough. We quantify how the recovery depends on (i) the radio channel noise, (ii) the optical signal-to-noise ratio (S/N), (iii) coherent radio residuals, and (iv) the spin temperature of the neutral gas.

The remainder of this paper is organized as follows. Section~\ref{sec:methods} describes the simulations, mock spectra, matched-filter search, Ly$\alpha$-dark mask, coherent radio residuals and catalogue matching. Section~\ref{sec:results} presents the radio-only recovery, the gain from the mask and its dependence on optical S/N, the false positives from coherent radio residuals, and the recovery in a heated IGM. Section~\ref{sec:discussion} discusses the radio and optical requirements and the limitations of the joint search, and Section~\ref{sec:conclusions} summarizes our conclusions. Throughout this work, we adopt a flat $\Lambda$CDM cosmology with $h=0.6751$, $\Omega_{\rm m}=0.3121$, and $\Omega_{\rm b}=0.04904$ \citep{Planck2020A&A...641A...6P}.

\section{Methods}
\label{sec:methods}

Our procedure has four steps. We first draw co-spatial Ly$\alpha$ and 21-cm sightlines from the same evolving density, ionization and temperature fields, and convert them into noisy mock optical and radio spectra. We then identify broad absorption candidates in the radio spectrum with a multiscale matched filter. Independently, we rebin the noisy Ly$\alpha$ spectrum onto the radio grid and mark channels whose optical flux is consistent with zero. Finally, we compare the radio-only and masked catalogues with the same one-to-one matching procedure, with and without coherent radio residuals added to the radio spectra.

\subsection{IGM simulations}

We take the density and ionization fields from the fiducial \textit{Meraxes} model of \citet{QinWyithe2026}. In a cosmological volume with a side length of $210\,h^{-1}\,{\rm cMpc}$, this model evolves galaxies along dark matter halo merger trees augmented down to the atomic-cooling threshold \citep{Balu2023MNRAS.520.3368B}. The galaxy ionizing emissivity is coupled to an inhomogeneous reionization calculation based on a customized version of {\tt 21cmFAST} \citep{Mesinger2011MNRAS.411..955M,Murray2020}. The density field is taken directly from the underlying $N$-body simulation, assuming that baryons follow the dark matter, and is discretized onto a $512^3$ grid with a cell size of $410\,h^{-1}\,{\rm ckpc}$. This resolution does not capture individual narrow absorption features. However, as our analysis targets broad troughs spanning at ${\gtrsim}10\,{\rm cMpc}$, this limitation is less critical. The ionization history of this model is calibrated against the observed distribution of Ly$\alpha$ effective optical depths from XQR-30+ over $5\lesssim z\lesssim6.1$ \citep{Bosman2022MNRAS.514...55B}. Models with different ionizing-source prescriptions distribute the escaping emissivity differently among galaxies, but once calibrated to these data they all predict a similarly late and extended reionization \citep{Qin2025PASA...42...49Q,QinWyithe2026}. We therefore use only the fiducial ionization lightcone.

The thermal state of the neutral IGM is less tightly constrained \citep{HERA2023ApJ...945..124H,Dhandha2025MNRAS.544.1608D,Nunhokee2025ApJ...989...57N,Kongprachaya2026arXiv260720371K,Soltinsky2026arXiv260715341S}. In our fiducial model, we turn off X-ray heating from star-forming galaxies, thereby maximizing the 21-cm absorption. In this limit, the simulated kinetic temperature ($T_{\rm K}$) is tightly coupled to the local overdensity, $\Delta\equiv1+\delta_{\rm b}$, where $\delta_{\rm b}$ is the baryon density contrast. We find that the kinetic temperature follows approximately $T_{\rm K}\propto\Delta^{0.6}$, close to the adiabatic expectation of $T_{\rm K}\propto\Delta^{2/3}$, as overdense gas is compressed and heated while underdense gas expands and cools. We assume strong coupling between the spin and kinetic temperatures, such that $T_{\rm S}\approx T_{\rm K}$. The volume-averaged temperatures are $T_{\rm S}\approx1.1$, 0.8 and $0.7\,{\rm K}$ at $z=7$, 6 and 5, respectively.

To determine how the abundance of broad troughs changes as the neutral IGM is heated, we also run models with a spatially uniform spin temperature\footnote{We verify that this approximation is reasonable using simulations that include X-ray heating from galaxies and span $10^{36}\le L_{\rm X}/{\rm SFR}[\,{\rm erg\,s^{-1}}\,M_\odot^{-1}\,{\rm yr}]\le10^{41}$ (see \citealt{Balu2023MNRAS.520.3368B} for details). Across this range, the spin temperature correlates only weakly with the local overdensity, with a scatter of ${\sim}0.1$\,dex at fixed overdensity.} of $1\le T_{\rm S}/{\rm K}\le20$. These uniform-$T_{\rm S}$ models keep the density and ionization fields of the fiducial model, but remove its temperature--density relation.

\subsection{Ly$\alpha$ and 21-cm transmission}
\label{sec:transmission}

\begin{figure*}
    \includegraphics[width=\textwidth]{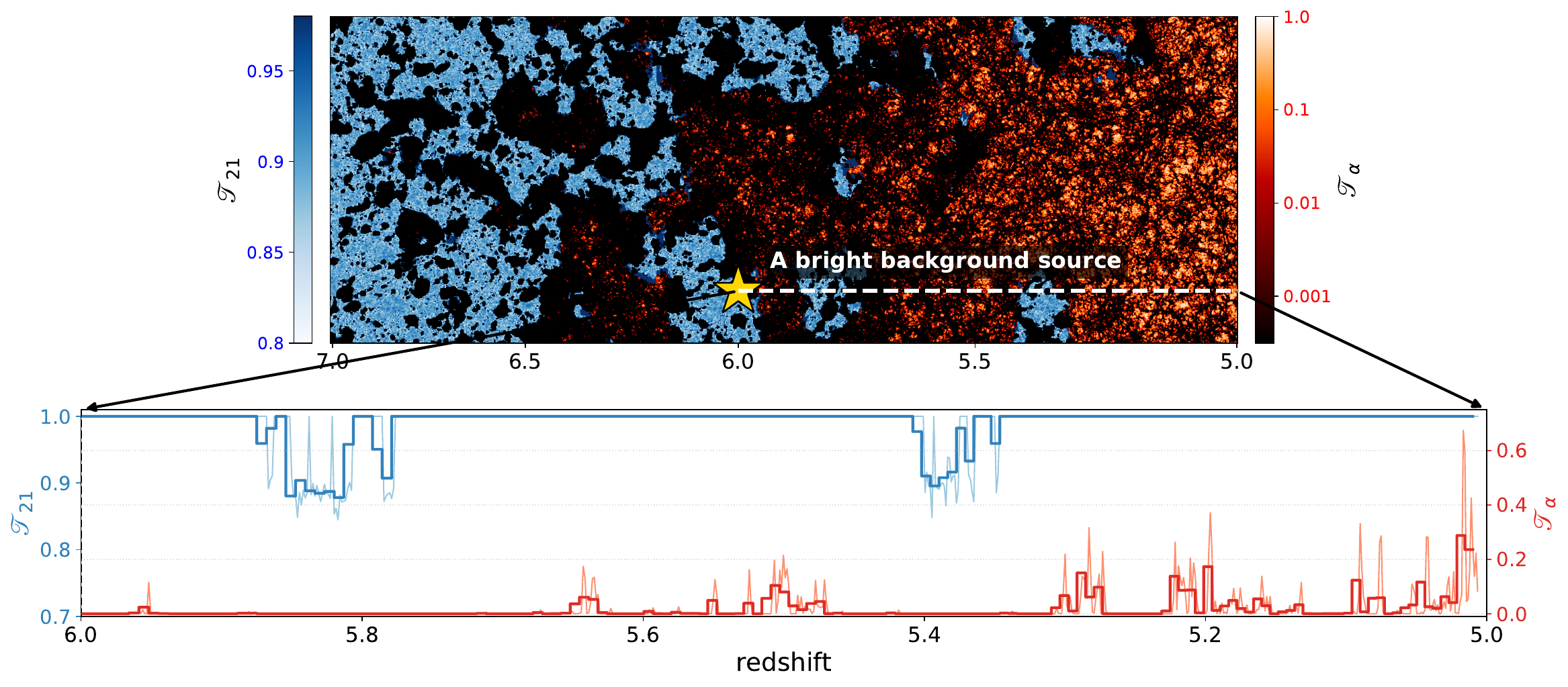}\vspace*{-5mm}
    \caption{\label{fig:transmission}\textit{Top:} Ly$\alpha$ and 21-cm transmission lightcones predicted by our model. The 21-cm transmission assumes no heating, so the absorption is maximized. \textit{Bottom:} a representative pair of sightlines at $z\lesssim6$ against a bright background source. The dark and light curves show the transmission at the simulation resolution and averaged over ${\sim}2$~cMpc, respectively. The extended Ly$\alpha$-dark region corresponds to a broad 21-cm trough.}
\end{figure*}

The Ly$\alpha$ transmission follows from the optical depth along each sightline. In ionized cells, the optical depth is evaluated with the fluctuating Gunn--Peterson (GP) approximation,
\begin{equation}
    \tau_{\alpha,{\rm GP}}
    =\left(\frac{3\pi\sigma_{\rm T}}{8}\right)^{1/2}
    c f_\alpha\lambda_{\alpha,0}H^{-1}
    n_{\rm H}x_{\rm HI,res},
    \label{eq:lya_optical_depth}
\end{equation}
where $\sigma_{\rm T}$ is the Thomson cross-section, $c$ is the speed of light, $f_\alpha$ and $\lambda_{\alpha,0}$ are the Ly$\alpha$ oscillator strength and rest-frame wavelength, $H(z)$ is the Hubble parameter at redshift $z$, $n_{\rm H}$ is the hydrogen number density, and $x_{\rm HI,res}$ is the residual neutral fraction. Neutral and partially ionized cells are opaque (i.e. $\mathscr{T}_\alpha=0$), while the transmitted fraction in ionized cells uses $\tau_{\alpha,{\rm GP}}$ directly: %
\begin{equation}
    \mathscr{T}_\alpha(\lambda)=\exp[-\tau_{\alpha,{\rm GP}}(\lambda)].
    \label{eq:lya_transmission}
\end{equation}

For 21-cm absorption, we calculate the optical depth in each cell as \citep{Furlanetto2006PhR...433..181F}
\begin{equation}
    \begin{aligned}
    \tau_{21}(\boldsymbol{x},z)
    ={}&27\times10^{-3}\,
    x_{\rm HI}(\boldsymbol{x},z)
    [1+\delta_{\rm b}(\boldsymbol{x},z)]    \left(\frac{1+z}{T_{\rm S}/{\rm K}}\right)
    \\
    &\times
    \left(\frac{\Omega_{\rm b}h^2}{0.023}\right)
    \left[
        \frac{1+z}{10}
        \frac{0.15}{\Omega_{\rm m}h^2}
    \right]^{1/2},
    \end{aligned}
    \label{eq:21cm_optical_depth}
\end{equation}
where $x_{\rm HI}$ is the neutral hydrogen fraction, and we ignore redshift-space distortions. The transmitted radio fraction is then
\begin{equation}
    \mathscr{T}_{21}(\nu)=\exp[-\tau_{21}(\nu)].
    \label{eq:21cm_transmission}
\end{equation}
Equation~(\ref{eq:21cm_optical_depth}) makes explicit that, for fixed density and ionization fields, the absorption strength scales as $T_{\rm S}^{-1}$.

\label{sec:simulation_fields}

Figure~\ref{fig:transmission} shows the resulting Ly$\alpha$ and 21-cm transmission lightcones between $z=7$ and 5, as well as an example sightline. The Ly$\alpha$ transmission is highly patchy, with extended dark regions corresponding to neutral islands. The 21-cm transmission shows broad troughs that align with these Ly$\alpha$-dark regions, demonstrating the potential of a joint search.

\subsection{Mock spectra, noise levels and true troughs}
\label{sec:mocks}
The simulation provides $512\times512$ sightlines, from which we construct the mock spectra. We restrict the analysis to $5 \le z \le 6$, a range that can be probed in whole or in part by dozens of known radio-loud sources at $z\gtrsim 5.5$ (e.g., \citealt{McGreer2006ApJ...652..157M,Willott2010AJ....139..906W,Belladitta2020A&A...635L...7B,Drouart2020PASA...37...26D,Banados2018ApJ...861L..14B,Banados2021ApJ...909...80B,Banados2025NatAs...9..293B} and this list\footnote{\url{https://tomassoltinsky.github.io//eor}}). We linearly interpolate the simulated fields onto an optical grid with a nominal sampling of $27\,{\rm km\,s^{-1}}$ and a radio grid with $\Delta\nu=0.390\,{\rm MHz}$. These grids are motivated by recent observations with X-shooter \citep{DOdorico2023MNRAS.523.1399D} and the upgraded Giant Metrewave Radio Telescope (uGMRT; \citealt{Kongprachaya2026arXiv260720371K}).

\begin{figure*}
    \includegraphics[width=0.92\textwidth]{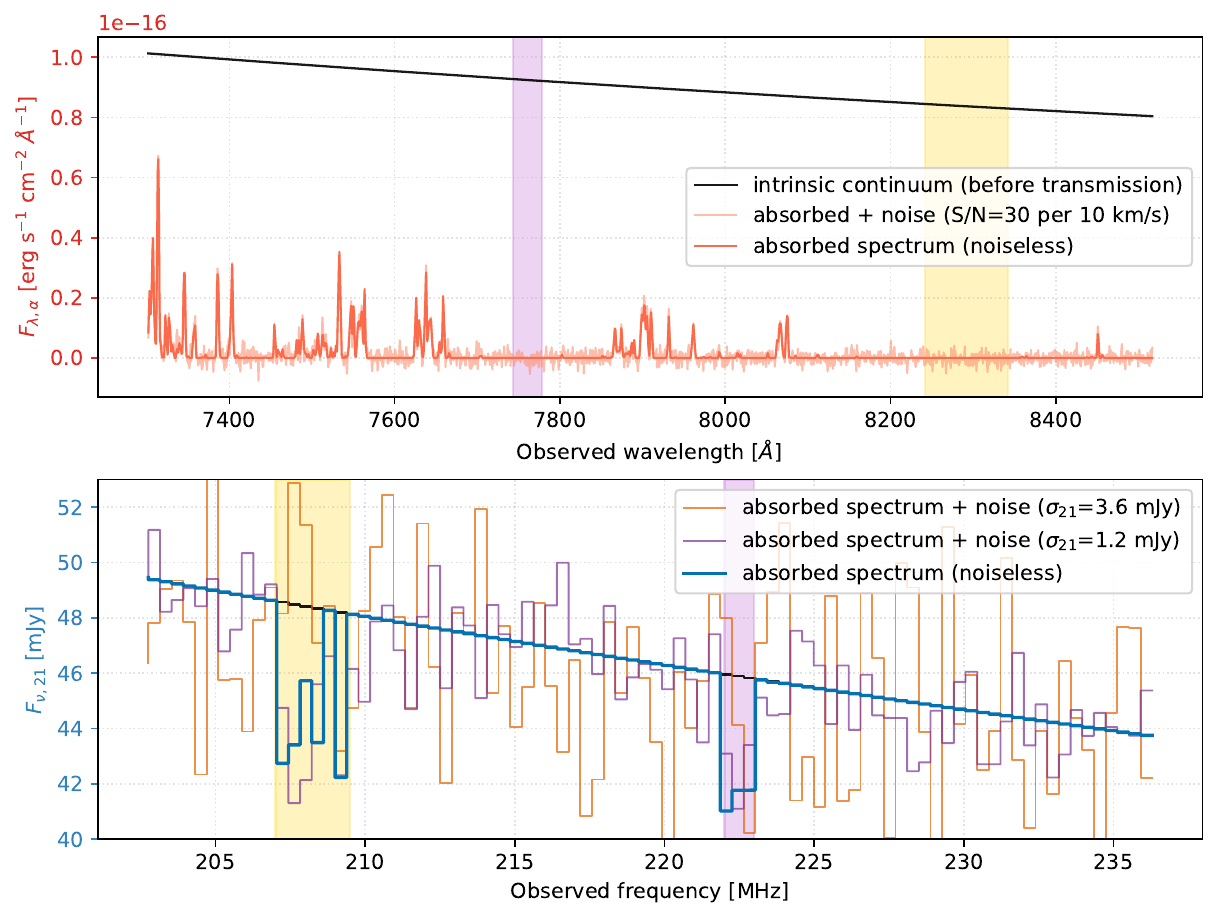}
    \caption{\label{fig:mock_spectra} Representative sightline with noiseless and noisy mock spectra. The top and bottom panels show the optical (Ly$\alpha$) and radio (21-cm) spectra, respectively. The yellow and purple bands highlight the two broad 21-cm troughs and the corresponding Ly$\alpha$-dark regions. The optical spectrum assumes an S/N of 30 per $10\ {\rm km\ s^{-1}}$ reference pixel. The radio panel shows representative noise realizations with RMS values of $\sigma_{21}=1.2$ and $3.6\ {\rm mJy}$ per channel. The troughs remain visually identifiable at $1.2\ {\rm mJy}$ but are largely obscured at $3.6{\rm mJy}$.}
\end{figure*}

We adopt power-law source continua, $F_\nu\propto\nu^\alpha$. Figure~\ref{fig:mock_spectra} shows an example continuum in the optical and the radio. The optical source has $F_\nu=0.2\,{\rm mJy}$ at $9000\,\text{\AA}$ and $\alpha_{\rm UV}=-0.5$, while the radio source has $F_\nu=50\,{\rm mJy}$ at $200\,{\rm MHz}$ and $\alpha_{\rm r}=-0.8$ (motivated by recent observations, e.g. PSO~J352$-$15 from \citealt{Kongprachaya2026arXiv260720371K} and the typical quasar continuum slope redward of Ly$\alpha$; see \citealt{VandenBerk2001AJ....122..549V}). We compute the absorbed flux for each channel and add independent Gaussian noise to every optical and radio channel of the noiseless spectra. Unless otherwise stated, we adopt an optical continuum S/N of 30 per $10\,{\rm km\,s^{-1}}$ pixel, motivated by the median S/N of the E-XQR-30 spectra \citep{DOdorico2023MNRAS.523.1399D}.

The appropriate radio noise is less clear, because current radio spectra have residuals that exceed expectations based on just the continuum and thermal noise. For an 18-hr uGMRT observation, the nominal thermal root-mean-square (RMS) noise is $\sigma_{21}{\sim}0.14\,{\rm mJy}$ per $0.390\,{\rm MHz}$ channel\footnote{This is estimated with natural weights using GMRT's online sensitivity calculator: \url{www.gmrt.ncra.tifr.res.in/gtac/etc/etc/rms/rms.html}.}, whereas the spectrum of PSO~J352$-$15 has a measured RMS of $3.6\,{\rm mJy}$ at the same resolution \citep{Kongprachaya2026arXiv260720371K}. A complementary analysis of the native $6.1\,{\rm kHz}$ channels reaches a similar RMS at this much finer resolution, likely because of differences in calibration and processing \citep{Soltinsky2026arXiv260715341S}. Even so, its noise is 2--3 times above the radiometer expectation. The origin of this excess scatter is uncertain and could include calibration or bandpass residuals that are not resolved at the final channel width. We therefore explore values of the radio RMS, $\sigma_{21}$, from 0.1 to $3.6\,{\rm mJy}$, extended down to $0.05\,{\rm mJy}$ for the heated-IGM tests of Section~\ref{sec:spin_temperature_results}. At $\sigma_{21}=3.6\,{\rm mJy}$, $\sigma_{21}$ should be read as an effective per-channel scatter rather than a purely thermal uncertainty. Coherent radio residuals on resolved frequency scales are treated separately (Section~\ref{sec:robustness_methods}).

We define a true 21-cm trough as at least two consecutive radio channels (i.e., $0.780\,{\rm MHz}$) with noiseless transmission $\mathscr{T}_{21}\le0.95$. At $z{\sim}6$, this frequency interval corresponds to a line-of-sight comoving scale of approximately $11\,{\rm cMpc}$. We define the trough centre as the absorption-weighted mean frequency, using weights $1-\mathscr{T}_{21}$, and place its boundaries half a channel beyond the centres of the outermost selected channels. We use these true troughs only to score the search. They do not depend on the radio noise, and they never enter the Ly$\alpha$-dark mask or the detected catalogues. For the spin-temperature test (Section~\ref{sec:spin_temperature_results}), we repeat the construction of true troughs with depth cuts $\mathscr{T}_{21}\le0.90$, 0.95, 0.97, 0.98, 0.99 and 0.995, keeping the minimum width of two radio channels.

Figure~\ref{fig:mock_spectra} also shows the noisy mock spectra of this sightline. The extended Ly$\alpha$-dark region remains apparent at an optical S/N of 30 per $10\,{\rm km\,s^{-1}}$ reference pixel. At the illustrative radio RMS of $\sigma_{21}=1.2\,{\rm mJy}$ per channel, the two broad 21-cm troughs remain visible by eye in this no-heating example, for which the absorption signal is maximized. At the measured scatter of $\sigma_{21}=3.6\,{\rm mJy}$ per channel discussed above, however, the channel-to-channel fluctuations become comparable to the trough depths, largely obscuring both features despite the unchanged underlying signal.

\subsection{Matched-filter radio search}
\label{sec:radio_filter}

\begin{figure*}
    \includegraphics[width=0.98\textwidth]{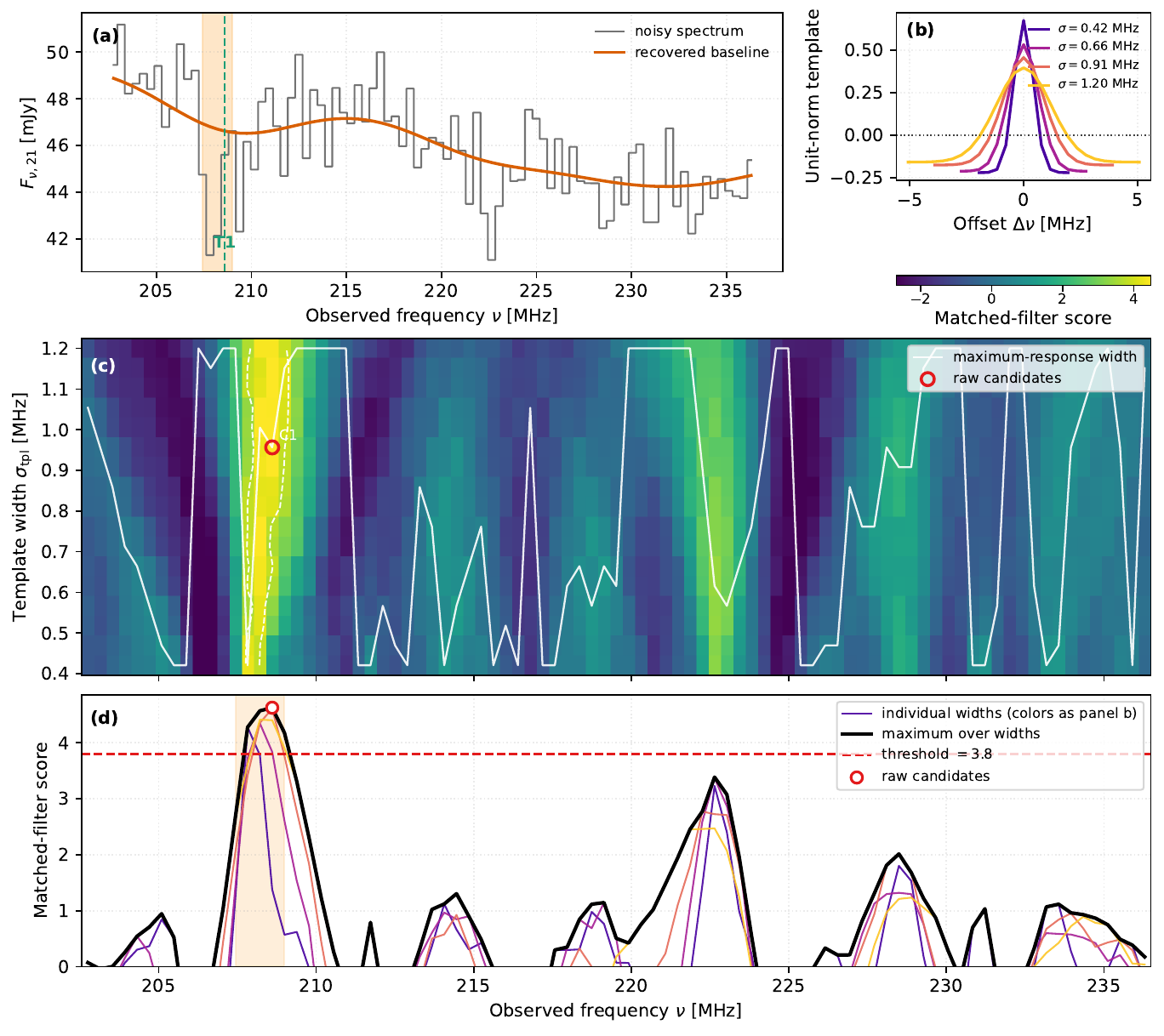}
    \caption{\label{fig:matched_filter_multiscale_candidates} Illustration of the multiscale matched-filter search for 21-cm absorption troughs along a representative noisy sightline. (a) Channelized 21-cm spectrum and recovered smooth baseline. The gold region and dashed green line indicate the final trough interval and centre. In this example, the noise level is reduced to $\sigma_{21}=1.2\,{\rm mJy}$ for illustration. (b) Representative zero-mean, unit-normalized Gaussian templates used in the search. (c) Matched-filter score as a function of observed frequency and template width \(\sigma_{\rm tpl}\). The solid white curve traces the template width producing the maximum score at each frequency, while the dashed white contour marks the detection threshold \(S=3.8\). Open red circles identify raw candidates. (d) Responses obtained with representative template widths and their pointwise maximum (black). Raw candidates are local maxima of this scale-maximized response that lie above the threshold and satisfy the minimum-separation criterion. Overlapping candidates are then merged into individual troughs.}
\end{figure*}

A matched filter cross-correlates the data with an expected signal profile and upweights fluctuations that remain coherent across the template width. It is therefore well suited to troughs that are individually weak in each channel but extend over several adjacent channels. Because the physical widths are not known in advance, we evaluate a bank of templates\footnote{The present template bank is deliberately agnostic about the population of trough widths and shapes. Simulation-calibrated priors on the joint distributions of 21-cm troughs and their corresponding Ly$\alpha$-dark regions could improve the search and enable a fully Bayesian analysis of the radio and optical spectra. We defer this extension to future work.} and retain the largest response at each frequency. Figure~\ref{fig:matched_filter_multiscale_candidates} illustrates this procedure.

We estimate the slowly varying radio continuum and bandpass with a Gaussian filter of width $\sigma_{\rm base}=3\,{\rm MHz}$, denoted by $B$, and define the positive absorption residual as
\begin{equation}
    R(\nu)=B[F_{21,{\rm obs}}](\nu)-F_{21,{\rm obs}}(\nu).
    \label{eq:radio_residual}
\end{equation}
In our controlled mock experiments, we normalize $R$ by the injected radio RMS, $\sigma_{21}$. In practice, one would need a noise estimate that is robust to absorption and to correlated bandpass residuals.

The template bank contains 24 Gaussian widths distributed linearly over $\sigma_{\rm tpl}=0.39$--$1.2\,{\rm MHz}$. Each discrete template is truncated at four standard deviations, its mean is subtracted, and it is normalized to unit squared norm. The scale-maximized statistic becomes
\begin{equation}
    S(j)=\max_s\sum_k \frac{R_k}{\sigma_{21}}g_s(j-k),
    \label{eq:matched_filter}
\end{equation}
where $g_s$ is the zero-mean template with width $s$. Radio-only candidates are local maxima with $S\ge3.8$, an empirical threshold that gives a reasonable balance between completeness and false positives in our mocks. We estimate their depth and full width at half maximum (FWHM) from a lightly smoothed ($\sigma=0.15$~MHz) residual, merge candidates with overlapping core intervals, and retain only detections with FWHM larger than 2 channels (i.e. FWHM$\ge0.78\,{\rm MHz}$). For the sightline of Fig.~\ref{fig:mock_spectra}, only one of the two broad troughs passes the threshold, and its recovered FWHM is less than half that of the true trough (Fig.~\ref{fig:matched_filter_multiscale_candidates}).

\subsection{Ly$\alpha$-dark mask from noisy optical data}
\label{sec:lya_prior}

Because the neutral gas that produces broad 21-cm troughs is also opaque to Ly$\alpha$ photons, we use the co-spatial optical spectrum to mark where 21-cm troughs can occur. We build this Ly$\alpha$-dark mask from the noisy mock optical spectrum alone, not from the noiseless simulated transmission. For each optical pixel $j$, we estimate the Ly$\alpha$ transmission and its uncertainty as
\begin{equation}
    \widehat T_{\alpha,j}
    =\frac{F_{\alpha,{\rm obs},j}}{F_{\alpha,{\rm cont},j}},
    \qquad
    \sigma_{T,j}
    =\frac{\sigma_{\alpha,j}}{F_{\alpha,{\rm cont},j}}.
    \label{eq:observed_lya_transmission}
\end{equation}
Here, $F_{\alpha,{\rm obs},j}$ and $\sigma_{\alpha,j}$ are the observed flux density and its RMS noise, respectively, while $F_{\alpha,{\rm cont},j}$ is the unabsorbed source continuum. In these controlled experiments, we use the continuum injected into the mock spectra, so continuum-recovery uncertainty is not included. This approximation is adequate for our purposes, because modern continuum-fitting methods recover the intrinsic continuum between Ly$\alpha$ and Ly$\beta$ with per cent level uncertainties (e.g. \citealt{Bosman2021MNRAS.503.2077B}).

A radio channel of width $0.390\,{\rm MHz}$ corresponds to a finite redshift interval. We map every optical pixel to its Ly$\alpha$ absorption redshift and collect all pixels within that interval into the set ${\cal J}_i$ for radio channel $i$. We then combine them into an inverse-variance weighted mean transmission,
\begin{equation}
    \overline T_{\alpha,i}
    =
    \sigma_{\overline T,i}^{2}{\displaystyle\sum_{j\in{\cal J}_i}
    \widehat T_{\alpha,j}\sigma_{T,j}^{-2}},
    \qquad
    \sigma_{\overline T,i}^{-2}
    =
    \sum_{j\in{\cal J}_i}\sigma_{T,j}^{-2}.
    \label{eq:binned_lya_transmission}
\end{equation}
Following the dark-pixel criterion of \citet{McGreer2015MNRAS.447..499M}, we classify a radio channel as Ly$\alpha$-dark when its rebinned optical flux is below twice the corresponding RMS noise. Because the flux and noise are divided by the same positive continuum, this can be written in transmission units as
\begin{equation}
    D_i=
    \begin{cases}
    1, & \overline T_{\alpha,i}<2\sigma_{\overline T,i},\\
    0, & \overline T_{\alpha,i}\ge2\sigma_{\overline T,i},
    \end{cases}
    \label{eq:dark_mask}
\end{equation}
where $D_i=1$ denotes a dark channel. The threshold is set by the noise of the optical spectrum rather than by a fixed transmission cut, so the mask can be built from observed data alone. Its selectivity varies with the optical S/N.


We compare two catalogues. The \emph{radio-only} catalogue uses $S\ge3.8$ throughout the band. The \emph{masked} catalogue applies the same radio threshold but retains only candidates whose centre channel has $D_i=1$. The optical spectrum only removes radio candidates that lie in transmissive regions, so it validates candidates but never replaces the radio detection statistic.

\subsection{Coherent radio residuals}
\label{sec:robustness_methods}

Observations of bright radio sources with the uGMRT \citep{Kongprachaya2026arXiv260720371K,Soltinsky2026arXiv260715341S} indicate that bandpass residuals are dominated by smooth structure on scales of a few MHz. They also show tentative evidence for shorter-period sinusoidal modes whose amplitudes exceed the thermal noise of the instrument. Such residuals can produce coherent spectral structure on the same frequency scales as broad 21-cm troughs. We quantify their effect on the recovery by adding a sinusoidal systematic residual,
\begin{equation}
    F_{21,{\rm sys}}(\nu_i)=A_{\rm sys}
    \operatorname{sinc}\!\left(\frac{\Delta\nu}{P_{\rm sys}}\right)
    \sin\!\left[
        \frac{2\pi(\nu_i-\nu_{\rm ref})}{P_{\rm sys}}+\phi
    \right],
    \label{eq:systematic}
\end{equation}
with peak amplitudes before channel averaging between 0 and $4\,{\rm mJy}$, and periods between $0.1$ and $24\,{\rm MHz}$. Here, $\nu_i$ is the centre of a radio channel of width $\Delta\nu=0.390\,{\rm MHz}$, and we define $\operatorname{sinc}(x)\equiv\sin(\pi x)/(\pi x)$. The sinc factor follows from averaging the sinusoid over a top-hat channel response. Modes with $P_{\rm sys}\gg\Delta\nu$ thus retain their original amplitude, whereas structure on scales shorter than the channel width is increasingly attenuated.

We fix $\nu_{\rm ref}$ to the central frequency of the radio band and set $\phi=0$ throughout, isolating the dependence on $A_{\rm sys}$ and $P_{\rm sys}$. We add the residual explicitly rather than including it in the thermal RMS used to normalize equation~(\ref{eq:matched_filter}), so that its effect on the matched-filter catalogue can be measured separately from that of thermal noise.

Unknown smooth radio residuals are also commonly handled by fitting and subtracting a low-order polynomial from the spectrum. We assess this approach by comparing the fiducial Gaussian baseline with blind full-band Legendre fits of degrees 2, 3, 5, 7, 9 and 12. These blind fits quantify the trade-off between suppressing smooth residuals and removing genuine broad absorption. They also provide a direct comparison with the independent Ly$\alpha$-dark mask.

\subsection{Catalogue matching and performance metrics}
\label{sec:metrics}

The recovery calculations draw 1000 sightlines without replacement and generate 10 noise realizations per sightline. We reuse the same sightlines and standardized Gaussian draws across noise levels and catalogue choices, so that differences between methods are paired rather than driven by different random samples. To map the dependence on optical data quality, we repeat the masked search for optical S/N values of 1, 2, 3, 5, 10, 20, 30 and 50 per $10\,{\rm km\,s^{-1}}$ pixel at each radio RMS, and add S/N${}=100$ for the heated-IGM tests of Section~\ref{sec:spin_temperature_results}.

We first form all compatible pairs of a true trough and a detection. A pair is compatible when the two frequency intervals overlap, or when their centres differ by at most one radio channel. Among these pairs, the Hungarian algorithm selects the global one-to-one assignment that minimizes the summed centre separation \citep{Kuhn1955Hungarian}. Each detection can match at most one true trough, and each true trough can receive credit from at most one detection. Unassigned detections are false positives, and unassigned true troughs are missed.

For pooled catalogues, we report the completeness and purity,
\begin{equation}
    {\cal C}=\frac{N_{\rm match}}{N_{\rm true}},
    \qquad
    {\cal P}=\frac{N_{\rm match}}{N_{\rm det}},
    \label{eq:metrics}
\end{equation}
together with the number of false positives, $N_{\rm det}-N_{\rm match}$. Here, $N_{\rm match}$, $N_{\rm true}$ and $N_{\rm det}$ are the total numbers of matched pairs, true troughs and detected candidates, respectively. We quote the false-positive rate as the mean number of false positives per spectrum, and summarize counting uncertainty with binomial 95 per cent Wilson intervals \citep{Wilson01061927}.

\section{Results}
\label{sec:results}

We first establish how well broad 21-cm troughs can be recovered from the radio spectra alone. We then use this radio-only catalogue as the reference to assess how much the Ly$\alpha$-dark mask reduces false positives, and how this gain depends on the optical S/N. Next, we quantify the false positives from coherent radio residuals, and finally the abundance and recovery of troughs in a heated IGM. Throughout this section, recovery is evaluated with the one-to-one matching described in Section~\ref{sec:metrics}.

\subsection{Recovery from the radio spectra alone}
\label{sec:radio_results}

\begin{figure}
    \centering
    \includegraphics[width=\columnwidth]{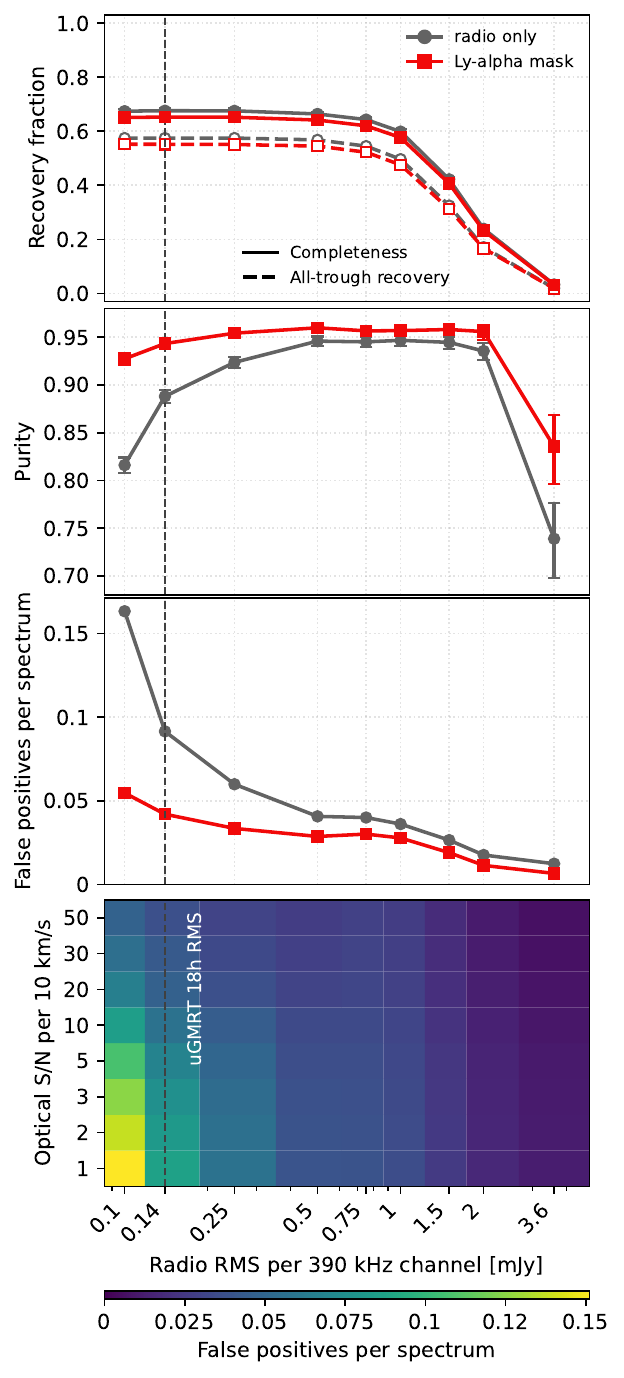}
    \caption{Recovery of broad 21-cm absorption troughs in the fiducial no-heating model, evaluated over 1000 sightlines and 10 noise realizations per sightline. The upper three panels compare the radio-only catalogue (grey circles) with the masked catalogue (red squares). Both use $S\ge3.8$ and ${\rm FWHM}\ge0.78\,{\rm MHz}$, but the latter keeps only candidates whose central channel satisfies the dark-pixel criterion of equation~(\ref{eq:dark_mask}). These panels adopt the default optical S/N of 30 per $10\,{\rm km\,s^{-1}}$ pixel. From top to bottom, they show the completeness (solid lines) together with the fraction of sightline realizations in which all true troughs are recovered (dashed lines), the purity, and the false-positive rate. The vertical dashed lines mark the nominal $0.14\,{\rm mJy}$ thermal RMS for an 18-hr uGMRT observation. The bottom panel shows the false-positive rate of the masked catalogue as a function of both optical S/N and radio RMS.}
    \label{fig:lya_guided_recovery}
\end{figure}

The radio-only curves in Fig.~\ref{fig:lya_guided_recovery} establish the baseline recovery. At $\sigma_{21}=0.1\,{\rm mJy}$, close to the nominal thermal RMS of an 18-hr uGMRT observation ($0.14\,{\rm mJy}$ per $0.390\,{\rm MHz}$ channel), the matched filter reaches a completeness of ${\sim}0.68$. The completeness declines only slowly up to $\sigma_{21}=1\,{\rm mJy}$ (${\sim}0.60$), then drops steeply, to only ${\sim}0.03$ at the measured $3.6\,{\rm mJy}$ scatter. At low noise, the completeness levels off instead of approaching unity. This suggests that the recovery there is limited not only by the noise, but also by the fixed detection threshold and by how the baseline subtraction and the template bank respond to the true troughs. The fraction of sightline realizations in which every true trough is recovered provides a more stringent measure. This fraction is ${\sim}0.57$ at 0.14--$0.25\,{\rm mJy}$ and falls with the completeness at higher noise.

The purity is ${\sim}0.82$ at $0.1\,{\rm mJy}$\footnote{The lower purity in the most sensitive bin is driven by an increase in the number of detections while the number of one-to-one matches has already saturated. At fixed spin temperature, equation~(\ref{eq:21cm_optical_depth}) implies that the trough depth increases with the neutral-gas density. Low-noise data therefore reveal shallower troughs from lower-density gas, and substructure near the boundaries of true troughs. Some of these features fail the requirement $\mathscr{T}_{21}\le0.95$ over at least $0.78\,{\rm MHz}$ and are counted as false positives, even though they trace genuine weaker absorption. The drop in purity at the lowest noise thus partly reflects our definition of a true trough, not only detections induced by noise.} and remains close to 0.95 between 0.5 and $1\,{\rm mJy}$. It stays above 0.9 from 0.25 to $2\,{\rm mJy}$, so as the spectra become noisier, the fixed $S\ge3.8$ threshold loses completeness rather than purity. At $3.6\,{\rm mJy}$, however, the purity also falls, to ${\sim}0.74$, because the false positives are no longer negligible compared with the few remaining true matches. The radio-only matched filter therefore yields a useful catalogue at low noise, but this is also where false positives are most numerous. Removing them is the job of the Ly$\alpha$-dark mask.

\subsection{Gain from Ly$\alpha$-dark validation}
\label{sec:guided_results}

The upper three panels of Fig.~\ref{fig:lya_guided_recovery} show one of the main results of this work. Applying the Ly$\alpha$-dark mask without changing the $S\ge3.8$ radio threshold leaves the completeness almost unchanged, while increasing the purity and reducing the false-positive rate. At $\sigma_{21}=0.1\,{\rm mJy}$, close to the nominal thermal RMS of $0.14\,{\rm mJy}$, the completeness changes only from ${\sim}0.68$ to ${\sim}0.65$. The purity, by contrast, rises from ${\sim}0.82$ to ${\sim}0.93$, and the false-positive rate falls from ${\sim}0.16$ to ${\sim}0.05$ per spectrum. The same holds at all radio RMS values tested. The mask rejects candidates in transmissive regions, but very few of those matched to true troughs.

The bottom panel of Fig.~\ref{fig:lya_guided_recovery} shows the joint dependence of the false-positive rate on the optical S/N and the radio RMS. The optical spectrum matters most when the radio data are most sensitive and the matched filter returns the largest candidate catalogue. At $\sigma_{21}=0.14\,{\rm mJy}$, the false-positive rate decreases from ${\sim}0.1$ per spectrum at S/N${}=1$ to ${\sim}0.04$ at the default S/N${}=30$. At $\sigma_{21}=1\,{\rm mJy}$, it depends only weakly on the optical S/N, and for $\sigma_{21}\ge2\,{\rm mJy}$ it stays below ${\sim}0.02$ for all optical S/N values considered. However, these noisy radio catalogues are already very incomplete (top panel), so their low false-positive rate is no advantage. Improving the optical spectrum is most valuable in the low-noise regime targeted by deeper radio observations.

\subsection{False positives from coherent radio residuals}
\label{sec:robustness_results}

\begin{figure*}
    \centering
    \includegraphics[width=0.95\textwidth]{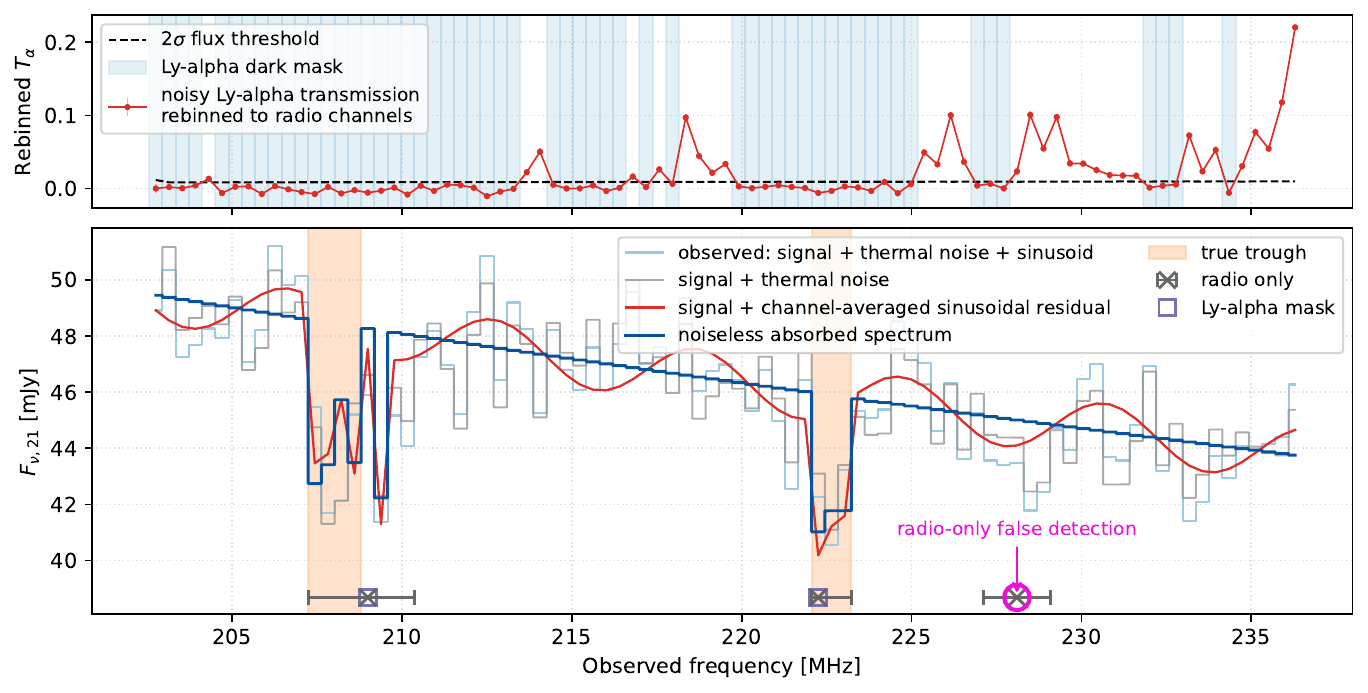}\vspace*{-2mm}
    \caption{Example co-spatial Ly$\alpha$ and 21-cm spectra with a sinusoidal radio residual, for an optical S/N of 30, $\sigma_{21}=1.2\,{\rm mJy}$, $A_{\rm sys}=1\,{\rm mJy}$ and $P_{\rm sys}=6\,{\rm MHz}$. The top panel shows the noisy Ly$\alpha$ transmission after inverse-variance binning onto the radio grid. The dashed curve marks twice the rebinned optical RMS, while the blue bands show channels whose measured flux lies below this threshold. The bottom panel separates the noiseless absorbed radio spectrum, thermal noise, sinusoidal residual and full mock observation. Orange bands mark true 21-cm troughs. Grey crosses mark the centres of the radio-only candidates, with the capped horizontal bars showing their recovered FWHM. Open purple squares identify the candidates whose retained by the Ly$\alpha$-dark mask, which filters the catalogue without changing the recovered centres or widths. The magenta circle highlights the radio-only false detection near $228\,{\rm MHz}$.}
    \label{fig:21cm_trough_recovery_example_systematics}
\end{figure*}

\begin{figure*}
    \centering
    \includegraphics[width=0.85\textwidth]{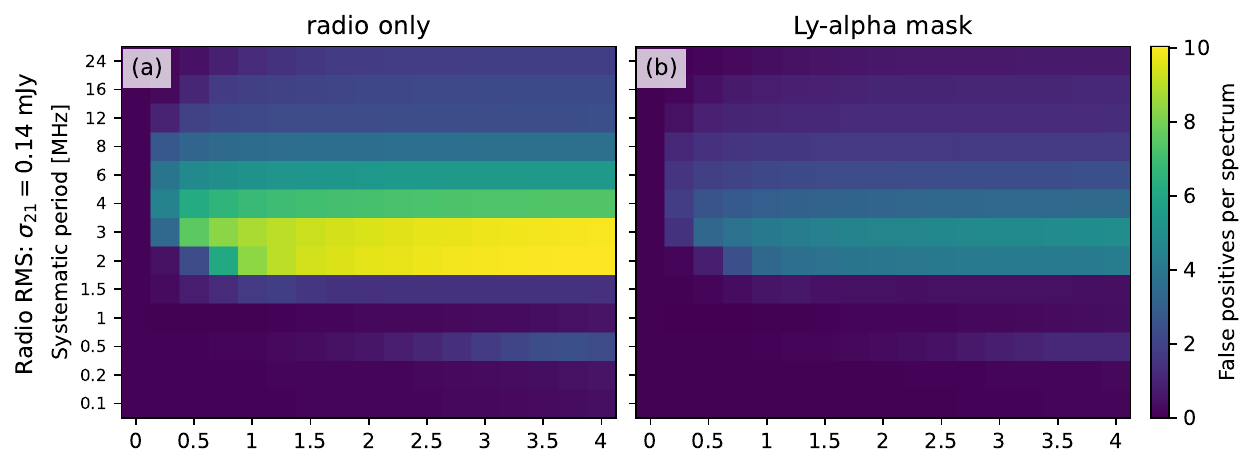}\\\vspace*{-2mm}
    \includegraphics[width=0.86\textwidth]{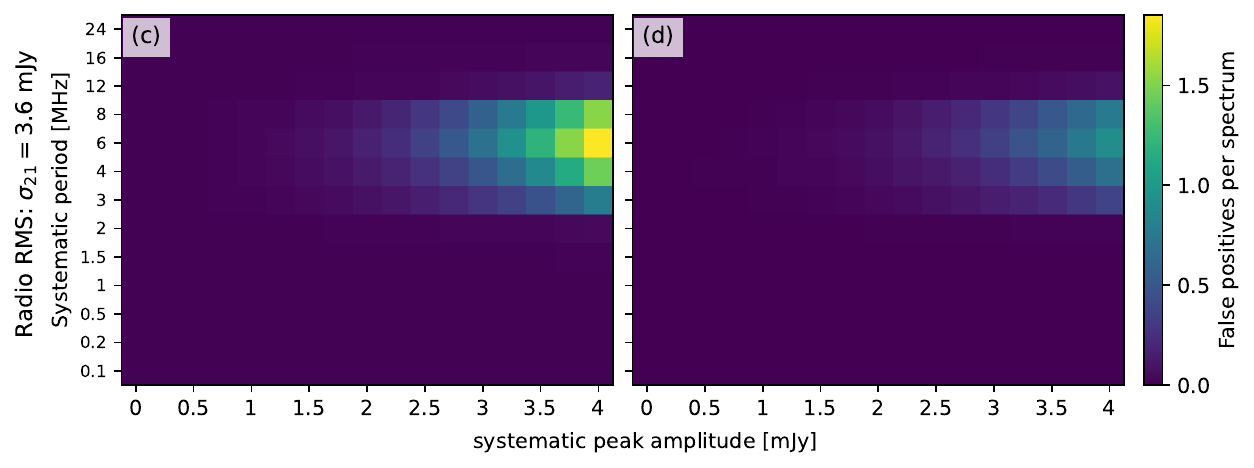}\vspace*{-3mm}
    \caption{Response of the trough search to sinusoidal radio residuals at two radio-noise levels (top: $\sigma_{21}=0.14\,{\rm mJy}$; bottom: $\sigma_{21}=3.6\,{\rm mJy}$). Each panel shows the false-positive rate as a function of the peak amplitude and period of the injected residual. The left panels show the radio-only catalogue, and the right panels the masked catalogue. Each row has its own colour scale.}
    \label{fig:systematics_heatmap}
\end{figure*}

Smooth systematic residuals can generate broad negative fluctuations that resemble 21-cm absorption after continuum removal. We now ask how many false positives such residuals create, and how many of them the Ly$\alpha$-dark mask rejects. 

Figure~\ref{fig:21cm_trough_recovery_example_systematics} shows an example sightline with a $1\,{\rm mJy}$ sinusoid of period $6\,{\rm MHz}$, added to radio data with $\sigma_{21}=1.2\,{\rm mJy}$. Its top panel shows the mask for this sightline, built as in Section~\ref{sec:lya_prior}. For this sightline, the radio-only search identifies the two true troughs near 209 and $222\,{\rm MHz}$, but also selects a fluctuation near $228\,{\rm MHz}$ produced by the combination of thermal noise and the sinusoidal residual. The recovered width of the lower-frequency candidate exceeds the true interval because smoothing blends the broad trough with the neighbouring narrow decrement. For the higher-frequency trough, the sinusoidal residual and thermal noise introduce an asymmetry that shifts the matched-filter maximum slightly away from the absorption-weighted true centre. Requiring the candidate centre to lie within the mask retains the candidates associated with the true troughs, while rejecting the spurious one. In this example, the radio spectrum supplies the candidates, and the Ly$\alpha$ spectrum tests whether each lies in highly opaque gas.

Figure~\ref{fig:systematics_heatmap} showcases one of the main results of this work.  It generalizes this comparison across sinusoidal amplitudes and periods, at the nominal thermal RMS of $0.14\,{\rm mJy}$ and at the measured scatter of $3.6\,{\rm mJy}$. At $0.14\,{\rm mJy}$ (panel a), the radio-only catalogue is most strongly affected by periods of ${\sim}2$--$4\,{\rm MHz}$, because a negative half-cycle then has a width comparable to that of the broad troughs targeted by the filter. The contamination peaks near $P_{\rm sys}=2$--$3\,{\rm MHz}$, where the false-positive rate approaches 10 per spectrum for large amplitudes. At longer periods, the $3\,{\rm MHz}$ Gaussian baseline removes an increasing fraction of the residual variation.  Short-period residuals are strongly suppressed at both noise levels. For $P_{\rm sys}\lesssim1\,{\rm MHz}$, the false-positive rate stays close to zero over most of the amplitude range, with only a modest increase for the largest amplitudes near $P_{\rm sys}=0.5\,{\rm MHz}$ at low noise. This is because channel averaging damps them, as described by the sinc factor in equation~(\ref{eq:systematic}), and because a rapidly oscillating residual averages out over a broad template. The damping is strongest for periods below the $0.390\,{\rm MHz}$ channel width, where a channel averages over one or more cycles.

The Ly$\alpha$-dark mask halves the false positives from the longer-period residuals that do mimic troughs. At $0.14\,{\rm mJy}$, it lowers the peak false-positive rate from ${\sim}10$ to 5 per spectrum (panels a and b), while retaining nearly all of the radio-only completeness similar to that shown in Fig.~\ref{fig:lya_guided_recovery}. The mask works because the optical spectrum is an independent tracer, whose instrumental residuals should be uncorrelated with those in the radio spectrum.

At $\sigma_{21}=3.6\,{\rm mJy}$ (panels c and d), the same residual amplitudes contribute less to the matched-filter significance, because the response is normalized by the larger radio RMS. The false-positive rate stays below ${\sim}2$ per spectrum and becomes appreciable mainly for $A_{\rm sys}\gtrsim2\,{\rm mJy}$ and $P_{\rm sys}\approx4$--$8\,{\rm MHz}$. The mask again halves its peak, from ${\sim}2$ to 1 per spectrum. As for thermal noise alone (Section~\ref{sec:guided_results}), this smaller contamination does not indicate a more favourable observing regime, because the radio-only completeness is only ${\sim}0.03$ at this noise level (Fig.~\ref{fig:lya_guided_recovery}).

\begin{figure}
    \centering
    \includegraphics[width=\columnwidth]{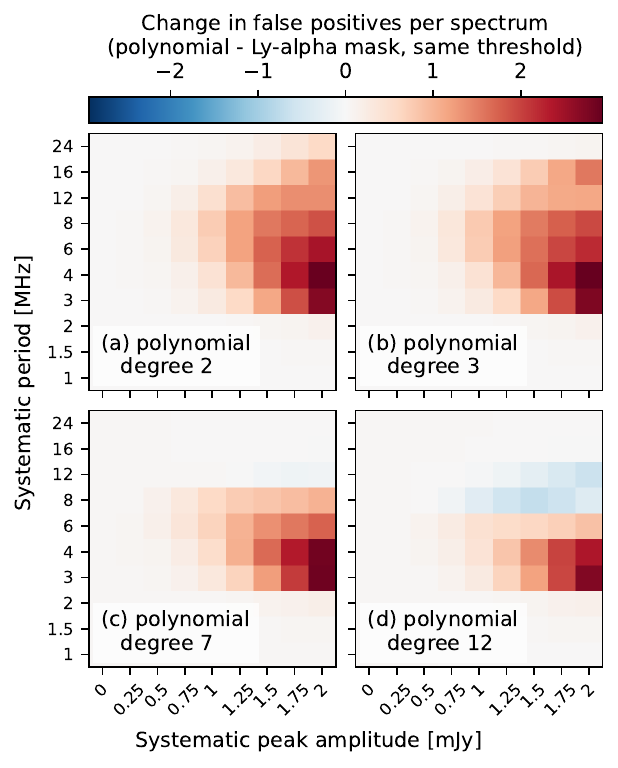}
    \caption{Change in the false-positive rate after subtracting blind full-band Legendre polynomials, at a fixed radio RMS of $\sigma_{21}=1.0\,{\rm mJy}$, relative to the fiducial Ly$\alpha$-masked search. The panels show polynomial degrees 2, 3, 7 and 12. Red cells indicate that polynomial subtraction produces more false positives than the Ly$\alpha$ mask, and blue cells fewer.}
    \label{fig:polynomial_systematics_heatmap}
    \includegraphics[width=\columnwidth]{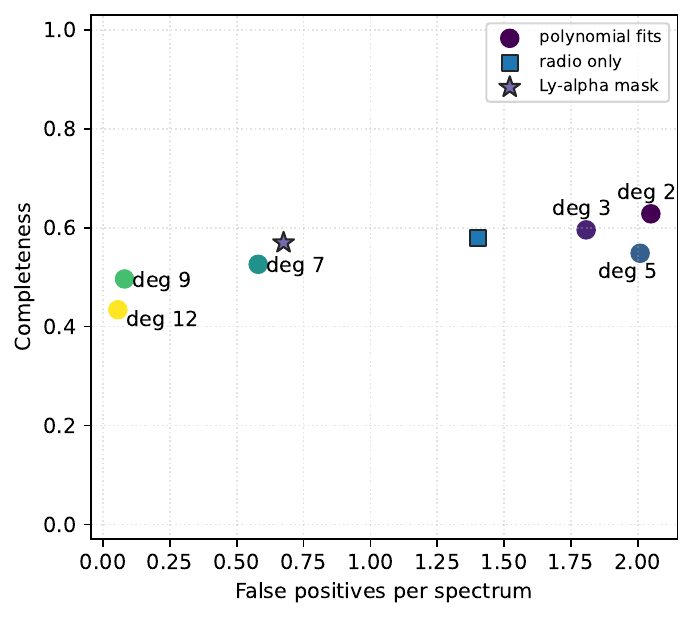}
    \caption{Completeness--contamination trade-off at $A_{\rm sys}=2\,{\rm mJy}$, $P_{\rm sys}=12\,{\rm MHz}$ and $\sigma_{21}=1\,{\rm mJy}$. Coloured circles show the radio-only results after fitting and subtracting blind full-band Legendre polynomials, each labelled by its polynomial degree. The square shows the fiducial radio-only catalogue search using the $3\,{\rm MHz}$ Gaussian baseline while the star shows the result obtained after applying the Ly$\alpha$ mask. The horizontal axis gives the false-positive rate, and the vertical axis the completeness after one-to-one matching.}
    \label{fig:polynomial_baseline_tradeoff}
\end{figure}

A common radio-only alternative to the mask is to fit and subtract a smooth polynomial when the form of the residuals is unknown (Section~\ref{sec:robustness_methods}). Figure~\ref{fig:polynomial_systematics_heatmap} evaluates this approach at a radio RMS of $1\,{\rm mJy}$ directly with the Ly$\alpha$-masked search. Where the sinusoidal residuals generate appreciable contamination, the degree-2, degree-3 and degree-7 fits generally leave more (up to 3) false positives than the mask, particularly for periods of ${\sim}3$--$6\,{\rm MHz}$ and large amplitudes. A degree-12 fit performs better for some longer-period residuals, but remains less effective than the mask at the shorter periods most likely to mimic the targeted troughs. 
No polynomial degree is best for all residual periods. Each degree suppresses false positives over some range of periods, but leaves them, or even adds to them, over another.

Figure~\ref{fig:polynomial_baseline_tradeoff} shows the associated cost in completeness for a representative case of $A_{\rm sys}=2\,{\rm mJy}$, $P_{\rm sys}=12\,{\rm MHz}$ and $\sigma_{21}=1\,{\rm mJy}$. Low degrees (2--5) keep a completeness similar to that of the fiducial Gaussian baseline (${\sim}0.58$), but leave more false positives than its ${\sim}1.4$ per spectrum. A degree-7 fit reduces the false-positive rate to ${\sim}0.6$ per spectrum with a completeness of ${\sim}0.52$, giving a trade-off close to that of the Ly$\alpha$ mask for this particular residual (see other cases in Fig.~\ref{fig:polynomial_systematics_heatmap}). Degrees 9 and 12 suppress the false positives to below 0.1 per spectrum, but also reduce the completeness, to ${\sim}0.43$ for degree 12. Increasing the polynomial degree thus removes more of the smooth residual, but eventually also fits out genuine broad absorption. The trend is not monotonic among the lower degrees, because their ability to represent the imposed sinusoid depends on the fitted scale. By contrast, the Ly$\alpha$-dark mask works differently. Because it tests each candidate against independent optical data, it removes false positives without changing the radio baseline, at a small cost in completeness. It does not replace careful radio calibration, but it makes the final catalogue less dependent on the choice of polynomial degree.

\subsection{Recovery in a heated IGM}
\label{sec:spin_temperature_results}

The no-heating model, in which the mean spin temperature of the neutral IGM is $\langle T_{\rm S}\rangle_{\rm HI} \approx$ 0.8 K at $z\approx6$,   maximizes the 21-cm optical depth, but is unlikely to describe the thermal state of the neutral IGM over the full redshift interval considered here. Recent 21-cm power-spectrum non-detections give model-dependent 95 per cent lower limits of $T_{\rm S}>2.5\,{\rm K}$ at $z=6.5$ from the Murchison Widefield Array (MWA) and $T_{\rm S}>15.6\,{\rm K}$ at $z=7.9$ from the Hydrogen Epoch of Reionization Array (HERA) \citep{Nunhokee2025ApJ...989...57N,HERA2023ApJ...945..124H}. These limits apply at higher redshifts than our $z=5$--6 lightcone and should not be imposed directly on the temperature of its remaining neutral islands. Nevertheless, if heating proceeds monotonically, they suggest that the neutral gas at $z=5$--6 is much warmer than in our no-heating model. We therefore ask how deep the radio data must be to recover troughs in such a heated IGM.

\begin{figure}
    \centering
    \includegraphics[width=\columnwidth]{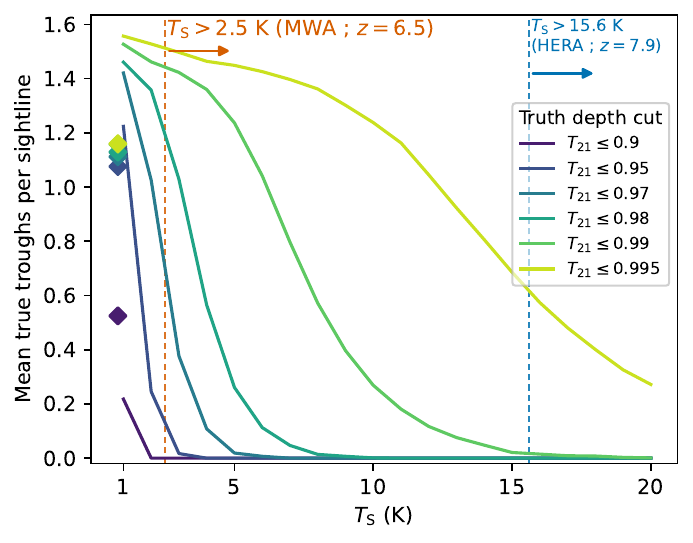}\vspace*{-5mm}
    \caption{Mean number of true 21-cm troughs per sightline as a function of a spatially uniform spin temperature. Coloured lines show depth cuts from $\mathscr{T}_{21}\le0.90$ to 0.995, while the diamonds show the no-heating model, placed at its volume-averaged temperature at $z=6$. The orange and blue dashed lines mark the model-dependent 95 per cent lower limits from the MWA at $z=6.5$ \citep{Nunhokee2025ApJ...989...57N} and HERA at $z=7.9$ \citep{HERA2023ApJ...945..124H}, respectively. These higher-redshift limits motivate the heated cases explored here, but are not imposed as priors.}
    \label{fig:trough_incidence_spin_temperature}
\end{figure}

Figure~\ref{fig:trough_incidence_spin_temperature} first shows how heating changes the population available to be detected. For every depth cut, the mean number of true troughs decreases as $T_{\rm S}$ increases, as expected from the $\tau_{21}\propto T_{\rm S}^{-1}$ scaling in equation~(\ref{eq:21cm_optical_depth}). For the fiducial depth cut, $\mathscr{T}_{21}\le0.95$, the incidence falls from ${\sim}1.2$ troughs per sightline at $T_{\rm S}=1\,{\rm K}$ to negligible values by $T_{\rm S}\approx3$--$4\,{\rm K}$. The diamonds show the no-heating model, whose temperature varies with density, so it has no single $T_{\rm S}$. It yields ${\sim}1.1$ fiducial troughs per sightline, slightly fewer than the uniform 1~K model, because adiabatic compression warms the overdense neutral gas, which would otherwise absorb most strongly.

The temperature dependence is weaker when shallower absorption counts as a true trough. For $\mathscr{T}_{21}\le0.99$, the incidence remains appreciable to ${\sim}10\,{\rm K}$, while the shallowest cut, $\mathscr{T}_{21}\le0.995$, yields ${\sim}0.6$ troughs per sightline at $T_{\rm S}=15.6\,{\rm K}$ and ${\sim}0.27$ at $20\,{\rm K}$. These two cuts correspond to decrements of only 1 and 0.5 per cent, and such troughs are harder to recover from noisy radio spectra than the fiducial ones. The number of troughs available to a search thus depends on both the IGM temperature and the shallowest absorption the observation can detect, which is set by the radio RMS and the detection threshold.

\begin{figure*}
    \centering
    \includegraphics[width=.98\textwidth]{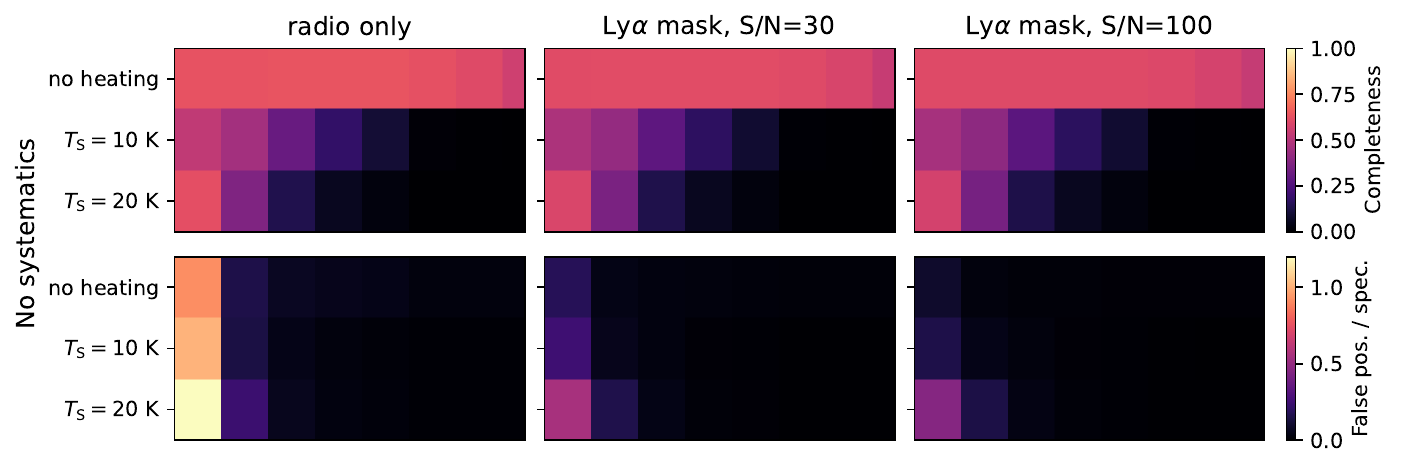}\\
    \includegraphics[width=.98\textwidth]{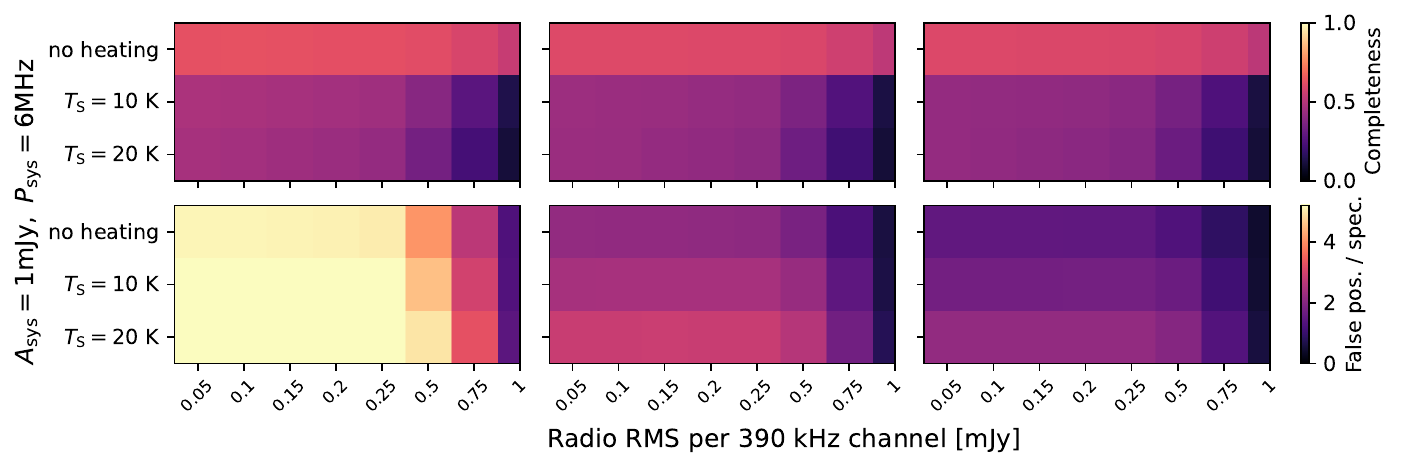}
    \caption{Recovery of shallow 21-cm troughs as a function of the radio RMS per $0.390\,{\rm MHz}$ channel, extended to $0.05\,{\rm mJy}$. True troughs use the depth cut $\mathscr{T}_{21}\le0.995$ and the minimum width of $0.78\,{\rm MHz}$. The upper block has no added radio residual, while the lower block includes a sinusoidal residual with $A_{\rm sys}=1\,{\rm mJy}$ and $P_{\rm sys}=6\,{\rm MHz}$. Within each block, the rows show the no-heating model and the uniform-$T_{\rm S}$ models with $T_{\rm S}=10$ and $20\,{\rm K}$. The upper panels show the completeness and the lower panels the false-positive rate. The three columns show the radio-only search and the masked search with an optical S/N of 30 and 100, all with the same $S\ge3.8$ threshold. Completeness is shown on a common colour scale from 0 to 1. The false-positive panels share a scale within each block, from 0 to 1.2 per spectrum without the residual and from 0 to 5 with it.}
    \label{fig:spin_temperature_recovery}
\end{figure*}

In Fig.~\ref{fig:spin_temperature_recovery} we show the recovery of these shallow troughs ($\mathscr{T}_{21}\le0.995$). The 10 and 20~K models lose completeness at a much lower radio RMS than the no-heating model. Their completeness is ${\sim}0.53$ (10~K) and ${\sim}0.62$ (20~K) in the lowest-noise bin, $0.05\,{\rm mJy}$, and becomes insignificant by $0.25\,{\rm mJy}$. The higher completeness at 20~K in the lowest-noise bin does not mean more detections, because there are far fewer true troughs at 20~K (Fig.~\ref{fig:trough_incidence_spin_temperature}). Heating therefore provides a double-punch to our detection changes: it leaves fewer troughs, and it makes those that survive harder to recover.

At $0.05\,{\rm mJy}$, the masked search removes most of the radio-only false positives in all three thermal models, with a completeness close to the radio-only value. Increasing the optical S/N from 30 to 100 suppresses the false-positive rate further, with only a small additional change in completeness. A representative sinusoidal residual ($A_{\rm sys}=1\,{\rm mJy}$, $P_{\rm sys}=6\,{\rm MHz}$) raises the radio-only false-positive rate to ${\sim}5$ per spectrum in the lowest-noise regime. The mask with S/N${}=30$ reduces this by a factor of ${\sim}2$, and the mask with S/N${}=100$ reduces it further. Detecting a representative population of troughs at $T_{\rm S}=10$--$20\,{\rm K}$ therefore requires a radio RMS of ${\sim}0.05$--$0.1\,{\rm mJy}$ and control of coherent radio residuals, with deeper optical spectra rejecting more of the remaining false positives.

\section{Discussion}
\label{sec:discussion}

Our results set two observational requirements: a radio RMS of ${\sim}0.05$--$0.1\,{\rm mJy}$ per channel if the neutral gas is heated (Section~\ref{sec:spin_temperature_results}), and optical spectra of the Ly$\alpha$ forest for the same sources. We discuss each in turn, followed by the limitations of our radio model.

The forthcoming low-frequency component of the Square Kilometre Array (SKA-Low) is the natural instrument to reach this radio RMS \citep{Ciardi2015aska.confE...6C,Cang2026aask.bookE.135C}. For example, \citet{Soltinsky2023MNRAS.519.3027S} adopted $A_{\rm eff}/T_{\rm sys}=600\,{\rm m^2\,K^{-1}}$ for SKA-Low when forecasting 21-cm absorption at $5\,{\rm kHz}$ resolution. Rescaling their thermal-noise estimate to our $0.390\,{\rm MHz}$ channels gives
\begin{equation}
    \sigma_{21,{\rm th}}{\sim} 0.039\,{\rm mJy}
    \sqrt{\left(\frac{10\,{\rm h}}{t_{\rm int}}\right)
    \left(\frac{0.390\,{\rm MHz}}{\Delta\nu}\right)}
    \left(\frac{600\,{\rm m^2\,K^{-1}}}{A_{\rm eff}/T_{\rm sys}}\right).
    \label{eq:ska_low_noise}
\end{equation}
Thus, the $0.05$--$0.1\,{\rm mJy}$ bins explored here are plausible thermal-noise targets for pointed SKA-Low observations, rather than purely illustrative limits. Whether a trough is detectable, however, also depends on the background source. All our recovery tests assume a radio continuum of $F_\nu=50\,{\rm mJy}$ at $200\,{\rm MHz}$. Because the absolute depth of an absorption feature scales linearly with the continuum flux density, the relevant sensitivity is approximately $\sigma_{21}/F_\nu$. At fixed noise, a fainter source thus needs a deeper fractional trough to reach the same matched-filter significance. The forecast therefore holds only if both long enough integrations and bright enough radio sources are available.

The same sources must also provide enough optical flux to measure the co-spatial Ly$\alpha$ forest. Future Ly$\alpha$ observations could consist of targeted optical or near-infrared spectroscopy of radio-selected high-redshift sources, with continuous coverage over the redshift interval searched in the radio spectrum. A higher optical S/N helps most at low radio noise (Section~\ref{sec:guided_results}), where it lowers the false-positive rate by a factor of ${\sim}3$ between S/N${}=1$ and 50 at $\sigma_{21}=0.14\,{\rm mJy}$. Our default S/N of 30 is representative of E-XQR-30 \citep{DOdorico2023MNRAS.523.1399D}, and spectra of lower S/N can still reject a useful fraction of the radio candidates. At an optical S/N of 100, currently achieved for a few XQR-30 quasars \citep{DOdorico2023MNRAS.523.1399D}, the mask further suppresses the remaining false positives in the heated-IGM tests, with little additional loss of completeness.

The mask classifies dark channels after rebinning onto the broad radio grid. Well-characterized binned noise, reliable continuum estimation, and control of sky residuals and of gaps in wavelength coverage thus matter more than high spectral resolution. The present mocks assume that the optical continuum and RMS are known exactly, so these effects should be included before forecasting a particular follow-up programme. Deeper spectra of known radio-bright quasars and future large-aperture spectroscopy of fainter optical counterparts offer complementary routes to increasing the target sample.

The radio side has its own limitations. Outside the sinusoidal tests, the channel scatter is modelled as white Gaussian noise, and the matched-filter normalization uses its known injected RMS. Real calibration residuals can also be non-Gaussian. The drop in purity at the lowest RMS partly reflects detections that are shorter or shallower than our definition of a true trough. This definition might need to be matched to the physical properties of the IGM and to the sensitivity of the observations. Finally, the Gaussian baseline can attenuate absorption on scales comparable to its $3\,{\rm MHz}$ width, and the Gaussian templates need not match the shapes of physical troughs. A natural next step is to describe the intrinsic radio continuum with a power law and fit its normalization and spectral index jointly with the trough parameters, accounting more directly for the continuum uncertainty. Establishing an appropriate continuum model and calibrating the resulting detection statistic are beyond these controlled tests, so we defer such a joint fit to future work.

\section{Conclusions}
\label{sec:conclusions}

Late reionization could leave extended neutral islands below $z{\sim}6$, which would imprint broad troughs in the 21-cm forest of radio-bright sources. We have developed a multiscale matched-filter search for these troughs, which correlates the baseline-subtracted radio spectrum with a bank of broad Gaussian templates. Following \citet{McGreer2015MNRAS.447..499M}, we validate its candidates with a mask of Ly$\alpha$-dark channels, in which the rebinned flux of a co-spatial, noisy Ly$\alpha$ spectrum is below twice its RMS noise. Our main results are the following.
\begin{enumerate}
    \item At $\sigma_{21}=0.1\,{\rm mJy}$, close to the nominal $0.14\,{\rm mJy}$ thermal RMS of an 18-hr uGMRT observation, the radio-only catalogue of the no-heating model has a completeness of ${\sim}0.68$ and a purity of ${\sim}0.82$. Applying the Ly$\alpha$-dark mask with the same radio threshold keeps a completeness of ${\sim}0.65$, while raising the purity to ${\sim}0.93$ and lowering the false-positive rate from ${\sim}0.16$ to ${\sim}0.05$ per spectrum.
    \item Coherent radio residuals can dominate the false positives. Sinusoids with periods of ${\sim}2$--$4\,{\rm MHz}$ produce up to ${\sim}10$ false positives per spectrum at $0.14\,{\rm mJy}$, and the Ly$\alpha$-dark mask halves this peak while retaining most of the radio-only recovery. Blind polynomial baselines either leave such false positives at some periods or, at high degree, remove genuine troughs. Unlike them, the mask uses information independent of the radio baseline, although it does not remove the need for radio calibration.
    \item Heating removes the fiducial troughs ($\mathscr{T}_{21}\le0.95$) by $T_{\rm S}\approx3$--$4\,{\rm K}$ and weakens those that remain. Shallow troughs ($\mathscr{T}_{21}\le0.995$) survive at $T_{\rm S}=10$--$20\,{\rm K}$, but recovering a representative population of them requires a radio RMS of ${\sim}0.05$--$0.1\,{\rm mJy}$. At $0.05\,{\rm mJy}$, the radio-only completeness is ${\sim}0.53$ (10~K) and ${\sim}0.62$ (20~K). Targeted SKA-Low observations could reach this sensitivity after binning to $0.390\,{\rm MHz}$. Raising the optical S/N from 30 to 100 rejects more false positives, with little further loss of completeness.
\end{enumerate}
These results show that matched filtering can recover a substantial fraction of broad 21-cm troughs once the required radio sensitivity is reached. The Ly$\alpha$-dark mask keeps most of that recovery while removing false positives with independent data, which matters most once continuum and bandpass residuals dominate over the thermal noise. Combining optical spectra with deep uGMRT or targeted SKA-Low observations could therefore yield reliable catalogues of the last neutral islands of reionization.

\section*{Data Availability}

The data underlying this article will be shared on reasonable request to the corresponding author. 

\section*{Acknowledgements}
The authors gratefully acknowledge the HPC support from OzSTAR and Gadi in Australia. YQ is supported by the ARC Discovery Early Career Researcher Award (DECRA) through fellowship \#DE240101129. OpenAI's ChatGPT and Codex were used to assist with language revision and the development and verification of the analysis workflow. Anthropic's Claude was used for some of the paper revision. All scientific decisions, validation of the results, and responsibility for the final manuscript remain with the authors.



\ifmsbibinclaude
  \bibliographystyle{../mnras}\bibliography{../references}
\else
  \bibliographystyle{mnras}\bibliography{references}
\fi



\bsp	
\label{lastpage}
\end{document}